\documentclass[aps,pra,reprint,nofootinbib,tbtags]{revtex4-1}
\usepackage{amsmath,amsfonts,amssymb,amsthm,bbm,color}
\usepackage[colorlinks=true,linkcolor=blue,citecolor=magenta,urlcolor=blue]{hyperref}
\usepackage[pdftex]{graphicx}
\renewcommand{\H}{\mathcal{H}}
\newcommand{\1}{\mathbbm{1}}
\newcommand{\bra}[1]{\langle #1|}
\newcommand{\ket}[1]{|#1\rangle}
\newcommand{\Tr}{\operatorname{tr}}
\usepackage[utf8]{inputenc}
\usepackage{lmodern}
\usepackage{hyperref}

\begin{document}
\title{Device-independent certification of two bits of randomness from \\ one entangled bit and Gisin's elegant Bell inequality}
\author{Ole Andersson}
	\email{ole.andersson@fysik.su.se}
\author{Piotr Badzi\c{a}g}
	\email{piotr.badziag@gmail.com}
\author{Irina Dumitru}
	\email{irina.dumitru@fysik.su.se}
\affiliation{Fysikum, Stockholms Universitet, S-106 91 Stockholm, Sweden}
\author{Ad\'{a}n Cabello}
	\email{adan@us.es}
\affiliation{Departamento de F\'{i}sica Aplicada II, Universidad de Sevilla, E-41012 Sevilla, Spain}

\begin{abstract}
We prove that as conjectured by Ac\'{\i}n {\em et al.}\ [Phys.\ Rev.\ A \textbf{93}, 040102(R) (2016)], two bits of randomness can be certified in a device-independent way from one bit of entanglement using the maximal quantum violation of Gisin's elegant Bell inequality. This suggests a surprising connection between
maximal entanglement, complete sets of mutually unbiased bases, and elements of symmetric informationally complete positive operator-valued measures, on one side, and the optimal way of certifying maximal randomness, on the other.
\end{abstract}

\maketitle

\section{Introduction}\vspace{-5pt}
Random numbers, i.e., numbers unpredictable to anyone, play a crucial role in cryptography, algorithms, and simulation. The possibility of certifying random numbers in a device-independent (DI) way, i.e., without making any assumption about the devices used to produce them and only assuming the impossibility of superluminal communication \cite{Colbeck06,PAMBMMOHLMM10,AM16}, is a great achievement of quantum information.

All methods for DI randomness certification \cite{Colbeck06,PAMBMMOHLMM10,AM16} require entangled pairs of systems and spacelike separated measurements whose outcomes violate one or several Bell inequalities \cite{Bell64} and, therefore, cannot be produced by any local realistic mechanism. The fact that entanglement and Bell inequality violation are the fundamental ingredients for DI randomness certification immediately raises two questions: (i) How many random bits can be certified from one ebit? (The {\em ebit} is the unit of bipartite entanglement and is defined as the amount of entanglement 
contained in a maximally entangled two-qubit state \cite{PR98}.) (ii) Which is the simplest Bell inequality, i.e., the one with the smallest number of settings, which allows for the DI certification of the maximal number of random bits? 
Question (i) has been answered recently. D'Ariano {\em et al.}\ \cite{DAriano2005} have proven that the maximum number of bits that can be certified in a DI way from one bit of entanglement using projective nondemolition or general demolition measurements is upper bounded by {\em two}, and Ac\'{\i}n {\em et al.}\ \cite{Acin2016} have proven analytically that this maximum can be {\em saturated} using a protocol based on a simultaneous maximal quantum violation of {\em three} Clauser-Horne-Shimony-Holt (CHSH) Bell inequalities \cite{CHSH69}. 
Question (ii) is still open. Intriguingly, Ac\'{\i}n {\em et al.}\ \cite{Acin2016} have also conjectured on the basis of numerical evidence that observing the maximum quantum violation of a single Bell inequality called ``the elegant Bell inequality'' (EBI) \cite{Gisin2009} is sufficient for the DI certification of two random bits. The fact that the EBI requires fewer settings than three CHSH Bell inequalities makes this conjecture interesting and worth trying to prove analytically.
In this paper, we provide such a proof.\vspace{-5pt}

\section{The elegant Bell inequality}\vspace{-5pt}
The EBI is a bipartite Bell inequality introduced by Gisin~\cite{Gisin2009} in which one of the parties, Alice, chooses among three dichotomic measurement settings, while the other party, Bob, chooses among four dichotomic measurement settings. If the possible outcomes are $\pm 1$ and $E_{k,l}$ denotes the mean value 
of the product of the outcomes of Alice's $k$th and Bob's $l$th settings, the EBI reads 
\begin{equation}\label{EBI}
\begin{split}
	S \equiv &\,E_{1,1}+E_{1,2}-E_{1,3}-E_{1,4}+E_{2,1}-E_{2,2}\\
	&+E_{2,3}-E_{2,4}+E_{3,1}-E_{3,2}-E_{3,3}+E_{3,4}\leq 6.
\end{split}
\end{equation}
Its maximum quantum violation is $S=4\sqrt{3}$ \cite{Acin2016}.

Besides the practical aspect that the EBI requires fewer settings than three CHSH Bell inequalities, there is also the exciting possibility that the answer to question (ii) would be the EBI. This would be remarkable. The adjective ``elegant'' in the EBI comes from the observation that its maximal quantum violation is achieved when Alice and Bob share an ebit, the eigenstates of Alice's three projective measurements form a complete set of three mutually unbiased bases (MUBs), and the eigenstates of Bob's four projective measurement can be divided into two sets, each of which defines a symmetric informationally complete positive operator-valued measure (SIC-POVM). MUBs and SIC-POVMs are two geometric structures of independent interest \cite{Wootters2006} and the fact that both might be simultaneously necessary for the optimal DI certification of maximal randomness from maximal entanglement would be quite surprising. 

Ac\'{i}n {\em et al.}\ \cite{Acin2016} have proposed a strategy for proving 
analytically that the EBI can be used for the DI certification of two random bits from one ebit. 
The strategy relies on the assumption that the maximal violation of the EBI is self-testing.
We have recently proven \cite{ABBDC17} that the maximal violation of the EBI is not self-testing
in the sense of Refs.\ \cite{McKague2010,McKague2010thesis}. 
However, the conjecture still holds and we prove it through a different strategy than 
the one proposed in Ref.~\cite{Acin2016}.\vspace{-3pt}

\section{Scenario}\vspace{-5pt}
We are interested in the following scenario.
Alice has a source of systems and a measurement device with four outcomes.
She uses them to perform a four-outcome measurement on each system produced by the source.
The generated outcomes are apparently unpredictable, i.e., after many measurements, Alice notices that the four outcomes 
appear with the same frequency and follow no pattern.
However, it might be that the outcomes are not so unpredictable as it seems and
someone else might be able to guess the outcomes of Alice's measurements.
That someone, whom we call the adversary, or Eve, could also be the manufacturer of Alice's device.
This means that the device is untrusted and that Alice is therefore interested in a device-independent certification of the randomness.
Here we propose two tests that Alice can perform to make sure that her device generates outputs which are completely 
unpredictable for everyone.
The tests, if passed, certify that the local guessing probability of Eve
does not exceed the minimal value $1/4$. If and only
if this is so, we say that Alice's measurement produces two random bits.\vspace{-10pt}

\section{Tests}\vspace{-5pt}
If we write $A_4$ for Alice's four-outcome POVM
and model Eve's substantiated guesses as outcomes $a$ of a local 
four-outcome POVM $F$
(if Eve measures $a$ she guesses that Alice measured $a$), 
the \emph{local guessing probability} of Eve is
\begin{equation}
	G=\max_{F}\sum_a P(a,a|A_4,F).
\end{equation}
The sum equals the probability that Eve makes a correct guess given that Alice measures $A_4$ and Eve measures $F$.
We maximize over all four-outcome POVMs that are local to Eve. The tests then certify that $G=1/4$.

\begin{figure}[t]
	\centering
	\includegraphics[width=8.8cm]{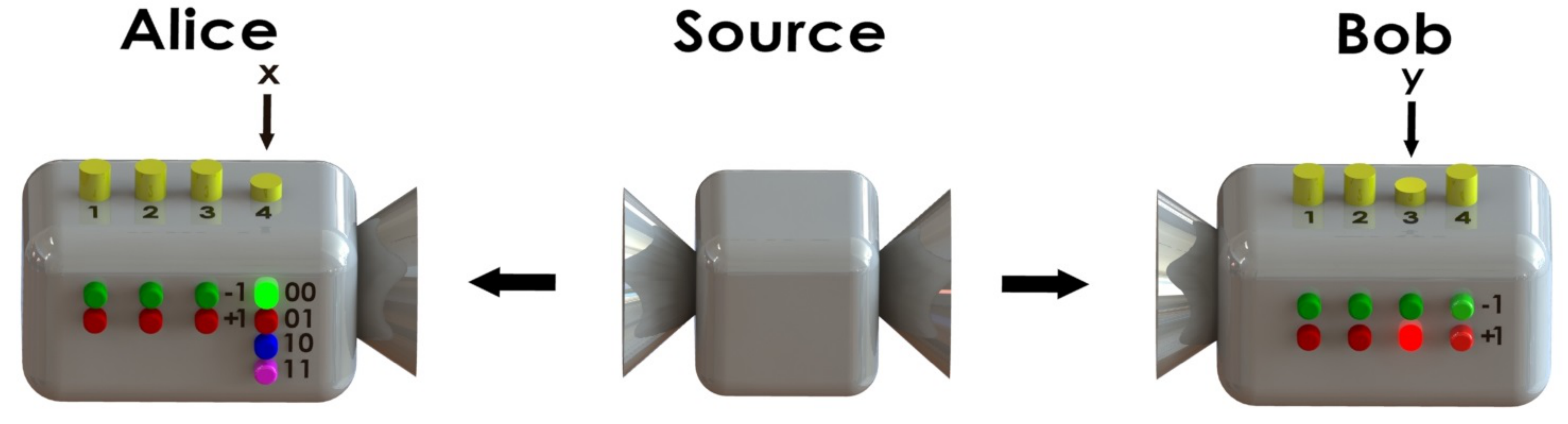}
	\caption{\label{Fig1}The source simultaneously emits two systems, one to each side. Buttons represent possible measurements. Light bulbs represent possible outcomes. 
	Alice and Bob wants to certify in a device-independent way that the two bits produced when Alice 
	presses her button $4$ are actually random (i.e., unpredictable even for an adversary who manufactured the devices).}
\end{figure}

The tests involve a third party, Alice's trusted friend Bob, who has access to a second system generated simultaneously 
by Alice's source. The scenario is sketched out in Fig.\ \ref{Fig1}.

For the tests, Alice needs three and Bob needs four measurement settings measuring local dichotomic observables.
We write $A_1,A_2,A_3$ and $B_1,B_2,B_3,B_4$ for Alice's and Bob's observables, respectively, and take their outcomes to be $-1$ and $+1$.
We also write $E_{k,l}$ for the expectation value of the products of the outcomes of Alice's $k$th and Bob's $l$th measurement
and $E_{a|k,l}$ for the expectation value of Bob's $l$th measurement
which is conditioned on the outcome of Alice's $k$th measurement, i.e.,
\begin{subequations}
\begin{align}
	E_{k,l} &= \sum_{a,b} ab\,P(a,b | A_k,B_l), \\
	E_{a|k,l} &= \sum_b b\, P(a,b|A_k,B_l).
\end{align}
\end{subequations}

\paragraph*{A test for the source.}
The first test is a Bell test.
To pass the test, Alice's and Bob's dichotomic measurements should
generate statistics indicating that the EBI is maximally violated: $S=4\sqrt{3}$.\vspace{3pt}

\paragraph*{A test for the measurement device.}
A necessary requirement for $G=1/4$ is that Alice's device generates an
apparently random output, i.e., $P(a|A_4)=1/4$ for all outcomes $a$.
We define a family of four qubit operators $Q=\{Q_a\}$ by
\begin{equation}\label{little a}
	Q_{a} = \gamma_a^0 \1 + \gamma_a^1 Z + \gamma_a^2 X + \gamma_a^3 Y,
\end{equation}
where $Z,X,Y$ are the Pauli operators and
\begin{subequations}\label{gammas}
	\begin{align}
	\gamma_a^0 &= P(a|A_4), \qquad & \\
	\gamma_a^1 &= \tfrac{\sqrt{3}}{2} (E_{a|4,1} + E_{a|4,2}), \\
	\gamma_a^2 &= \tfrac{\sqrt{3}}{2} (E_{a|4,1} + E_{a|4,3}), \\
	\gamma_a^3 &= -\tfrac{\sqrt{3}}{2} (E_{a|4,2} + E_{a|4,3}). 
	\end{align}
\end{subequations}
The second test is passed if $P(a|A_4)=1/4$ and $Q$ is an 
\emph{extremal} four-outcome qubit POVM. Here Bob uses the same three observables $B_1,B_2,B_3$ 
used in the first test.
Below we describe how to determine that $Q$ is an extremal POVM.\vspace{3pt}

Since the tests only require an analysis of the measurement statistics and assume nothing about either the 
devices used to generate this statistics or the measurement device used by Eve, they ensure 
that the randomness generated by Alice is genuine and device-independent.

The simplest scenario that passes the two tests is the following.
Suppose that Alice and Bob share two qubits in the singlet state,
\begin{equation}\label{singlet}
	\ket{\phi_+}=\tfrac{1}{\sqrt{2}}(\ket{00}+\ket{11}).
\end{equation} 
If Alice measures three dichotomic observables which correspond to the Pauli observables
\begin{equation}
	A_1 = Z, \qquad A_2 = X, \qquad A_3 = Y,
\end{equation}
and Bob measures four observables which correspond to 
\begin{subequations}
\begin{align}
	B_1 &= \tfrac{1}{\sqrt{3}}(Z+X-Y), & B_3 &= \tfrac{1}{\sqrt{3}}(-Z+X+Y),\\
	B_2 &= \tfrac{1}{\sqrt{3}}(Z-X+Y), & B_4 &= \tfrac{1}{\sqrt{3}}(-Z-X-Y),
\end{align}
\end{subequations}
then the EBI is maximally violated, which means the first test is passed.
Furthermore, if Alice measures the four-outcome POVM $A_4$ whose elements
correspond to the four linearly independent unit rank projectors
\begin{subequations}
\begin{align}
	A_{1|4} &= \tfrac{1}{4}\big(\1-\tfrac{1}{\sqrt{3}}(Z+X+Y)\big),\\
	A_{2|4} &= \tfrac{1}{4}\big(\1-\tfrac{1}{\sqrt{3}}(Z-X-Y)\big),\\
	A_{3|4} &= \tfrac{1}{4}\big(\1+\tfrac{1}{\sqrt{3}}(Z-X+Y)\big),\\
	A_{4|4} &= \tfrac{1}{4}\big(\1+\tfrac{1}{\sqrt{3}}(Z+X-Y)\big),
\end{align}
\end{subequations}
then $Q$ defined by Eq.\ \eqref{little a} equals $A_4$,
which is extremal according to  
the discussion in Sec.\ \ref{ex qu povms}.
The requirement $P(a|A_4)=1/4$ is also satisfied and,
hence, the second test is also fulfilled.\vspace{-14pt}

\section{Proof}\vspace{-5pt}
We now prove that for any quantum state $\ket{\psi}$ generated by Alice's source and shared with Bob and Eve,
and for any $A_1,A_2,A_3,A_4$ local to Alice, $B_1,B_2,B_3,B_4$ local to Bob, and $F$ local to Eve,
if the two tests have been passed, then $\sum_{a} P(a,a|A_4,F)=1/4$ and therefore $G=1/4$.

In Ref.\ \cite{ABBDC17}, we have shown that a maximal violation of the EBI implies the existence of an isometry
$\Phi=\Phi_A\otimes\Phi_B\otimes\1_E$, 
\begin{equation}
\begin{split}
	\hspace{-5.7pt}\Phi:\H_A \otimes \H_B \otimes \H_E 
	&\to (\H_A \otimes \H_2) \otimes (\H_B\otimes \H_2) \otimes \H_E \\
	=&\, (\H_A \otimes \H_B \otimes \H_E ) \otimes (\H_2 \otimes \H_2),
	\end{split}
\end{equation}
such that $\Phi(\ket{\psi})=\ket{\chi}\otimes \ket{\phi_+}$ for some $\ket{\chi}$ in $\H_A\otimes \H_B\otimes \H_E$ and such that
\begin{subequations}
\begin{align}
	\Phi(B_1\ket{\psi}) &= \tfrac{1}{\sqrt{3}} \Big(\ket{\chi}\otimes\big(\1\otimes(Z+X)\ket{\phi_+}\big) \notag  \\
						&{}\hspace{68pt} - J\ket{\chi}\otimes(\1\otimes Y\ket{\phi_+}) \Big), \\
	\Phi(B_2\ket{\psi}) &= \tfrac{1}{\sqrt{3}} \Big(\ket{\chi}\otimes\big(\1\otimes(Z-X)\ket{\phi_+}\big) \notag   \\
						&{}\hspace{68pt} + J\ket{\chi}\otimes(\1\otimes Y\ket{\phi_+}) \Big),  \\
	\Phi(B_3\ket{\psi}) &= \tfrac{1}{\sqrt{3}} \Big(\ket{\chi}\otimes\big(\1\otimes(-Z+X)\ket{\phi_+}\big) \notag   \\
						&{}\hspace{68pt} + J\ket{\chi}\otimes(\1\otimes Y\ket{\phi_+}) \Big),  \\
	\Phi(B_4\ket{\psi}) &= \tfrac{1}{\sqrt{3}} \Big(\ket{\chi}\otimes\big(\1\otimes(-Z-X)\ket{\phi_+}\big) \notag   \\
						&{}\hspace{68pt} - J\ket{\chi}\otimes(\1\otimes Y\ket{\phi_+})  \Big). 
\end{align}
\end{subequations}
Here, $\H_A$, $\H_B$, and $\H_E$ are the Hilbert spaces of Alice, Bob, and Eve, $\H_2$ is a two-dimensional Hilbert space with a computational basis $\{\ket{0},\ket{1}\}$, the state $\ket{\phi_+}$ is the two-qubit singlet state
defined in Eq.~\eqref{singlet},
and $J$ is an involution (i.e., $J^2$ is the identity) on the support of $\Phi_B\otimes\1_E$ which commutes with every operator local to~Eve.

On the support of $\Phi_A$, each $A_{a|4}$, i.e., the element of $A_4$ corresponding to outcome $a$, can be represented by an operator $R_a$ acting on $\H_A\otimes\H_2$.
If we expand $R_a$ as
\begin{equation}
R_a=R_a^0\otimes \1 + R_a^1\otimes Z + R_a^2\otimes X + R_a^3\otimes Y,
\end{equation}
where each $R_a^k$ is a Hermitian operator on $\H_A$, then
\begin{subequations}\label{alphas}
	\begin{align}
	\gamma_a^0 &\equiv \bra{\psi} A_{a|4} \ket{\psi} = \bra{\chi} R_a^0 \ket{\chi}, \\
	\gamma_a^1 &\equiv \tfrac{\sqrt{3}}{2} \bra{\psi} A_{a|4} (B_1 + B_2) \ket{\psi} = \bra{\chi} R_a^1 \ket{\chi},	\\
	\gamma_a^2 &\equiv \tfrac{\sqrt{3}}{2} \bra{\psi} A_{a|4} (B_1 + B_3) \ket{\psi} = \bra{\chi} R_a^2 \ket{\chi}, \\
	\gamma_a^3 &\equiv -\tfrac{\sqrt{3}}{2} \bra{\psi} A_{a|4} (B_2 + B_3) \ket{\psi} = \bra{\chi} R_a^3 J \ket{\chi}.
	\end{align}
\end{subequations}
The family of operators $Q=\{Q_a\}$ on $\H_2$ defined by
\begin{equation}\label{little aa}
Q_a = \gamma_a^0 \1 + \gamma_a^1 Z + \gamma_a^2 X + \gamma_a^3 Y
\end{equation}
forms an extremal four-outcome POVM by the second test.

The operator $J$ is diagonalizable with eigenvalues $-1$ and $+1$. 
We write $J_{\pm}$ for the orthogonal projections onto its $\pm 1$ eigenspaces.
Also, inspired by Ac\'{i}n \emph{et al.}, we define normalized states $\ket{\varphi_{\pm,a}}$ by 
\begin{equation}
	\ket{\varphi_{\pm,a}} = J_{\pm}F_a\ket{\chi}/\sqrt{q_{\pm,a}}.
\end{equation}
Then,
\begin{equation}\label{first}
\begin{split}
\hspace{-5pt}\gamma_a^k
&=\sum_{a'} \bra{\chi} F_{a'} J_+ R_a^k J_+ F_{a'}\ket{\chi} + \bra{\chi} F_{a'} J_- R_a^k J_- F_{a'}\ket{\chi} \\
&=\sum_{a'} q_{+,a'} \bra{\varphi_{+,a'}} R_a^k \ket{\varphi_{+,a'}} + q_{-,a'} \bra{\varphi_{-,a'}} R_a^k \ket{\varphi_{-,a'}} \\
&\equiv\sum_{a'} q_{+,a'} \beta_a^{k;+,a'} + q_{-,a'} \beta_a^{k;-,a'}
\end{split}
\end{equation}
for $k=0,1,2$ and
\begin{equation}\label{second}
\begin{split}
\hspace{-5pt}\gamma_a^3
&=\sum_{a'} \bra{\chi} F_{a'} J_+ R_a^3 J_+ F_{a'} \ket{\chi} - \bra{\chi} F_{a'} J_- R_a^3 J_- F_{a'} \ket{\chi} \\
&=\sum_{a'} q_{+,a'} \bra{\varphi_{+,a'}} R_a^3 \ket{\varphi_{+,a'}} - q_{-,a'} \bra{\varphi_{-,a'}} R_a^3 \ket{\varphi_{-,a'}} \\
&\equiv\sum_{a'} q_{+,a'} \beta_a^{3;+,a'} - q_{-,a'} \beta_a^{3;-,a'}.
\end{split}
\end{equation}
Here we have, without loss of generality, assumed that $F$ is projective.
Next, define four-outcome qubit POVMs $R^{\pm,a'}=\{R^{\pm,a'}_a\}$ as
\begin{subequations}\label{theRs}
	\begin{align}
	R^{+,a'}_a\!&=\beta_a^{0;+,a'} \1 + \beta_a^{1;+,a'} 
	Z + \beta_a^{2;+,a'} X + \beta_a^{3;+,a'} Y, \\
	R^{-,a'}_a\!&=\beta_a^{0;-,a'} \1 + \beta_a^{1;-,a'}
	Z + \beta_a^{2;-,a'} X - \beta_a^{3;-,a'} Y.
	\end{align}
\end{subequations}
From Eqs.\ \eqref{first} and \eqref{second} follow that $Q_{a} = \sum_{\pm,a'} q_{\pm,a'} R_a^{\pm,a'}$, which is 
a convex decomposition of $Q$. Since $Q$ is extremal, $R_a^{\pm,a'}=Q_{a}$ and, hence, $\beta_a^{k;\pm,a'}=\gamma_{a}^k$ for all $a'$. 
In particular, $\beta_a^{0;\pm,a}=\gamma_a^0=1/4$ for all $a$.
Now,
\begin{equation}
\begin{split}
\hspace{-5pt}\sum_{a} P&(a,a|A_4,F)=\sum_{a} \bra{\psi} A_{a|4} F_a \ket{\psi} \\
&= \sum_{a} \bra{\chi} R_a^0 F_a\ket{\chi} \\
&= \sum_{a} \bra{\chi} F_a J_+ R_a^0 J_+ F_a \ket{\chi} 
+ \bra{\chi} F_a J_- R_a^0 J_- F_a \ket{\chi} \\
&= \sum_{a} q_{+,a} \beta_a^{0;+,a} + q_{-,a} \beta_a^{0;-,a} \\
&= 1/4.
\end{split}
\end{equation}
Since we have not assumed anything about Eve's measurement, this proves that $G=1/4$.

\section{Extremal qubit POVMs}\label{ex qu povms}\vspace{-6pt}
POVMs of a fixed number of outcomes form a convex set. Its extremal elements are those that cannot be written as nontrivial convex combinations of other POVMs.
D'Ariano \emph{et al.}\ \cite{DAriano2005} have classified all extremal POVMs with discrete output sets.
According to this classification, a four-outcome qubit POVM is extremal if, and only if, it consists of four linearly independent one-dimensional 
projectors. 
The elements of $Q$ defined by Eq.\ \eqref{little a} are one-dimensional projectors provided that $\Tr Q_a>0$ and $\det Q_a=0$.
The former condition is satisfied if $P(a|A_4)>0$ and the latter condition is satisfied if
\begin{equation}
\begin{split}
\big( E_{a|4,1} + E_{a|4,2} \big)^2
&+ \big( E_{a|4,1} + E_{a|4,3} \big)^2 \\
&+ \big( E_{a|4,2} + E_{a|4,3} \big)^2
= \tfrac{4}{3}P(a|A_4)^2
\end{split}
\end{equation}
for all $a$. Moreover, the projectors are linearly independent provided
the vectors $[\gamma^{0}_{a}\,\gamma^{1}_{a}\,\gamma^{2}_{a}\,\gamma^{3}_{a}]^T$ are linearly independent, 
where the $\gamma^{k}_{a}$s are defined as in Eq.\ \eqref{gammas}. Given that $\gamma^{0}_{a}=P(a|A_4)=1/4$ for all $a$,
this is equivalent to the condition that the matrix of conditional expectation values, 
\begin{equation}
\begin{bmatrix}
E_{1|4,1} & E_{1|4,2} & E_{1|4,3} \\
E_{2|4,1} & E_{2|4,2} & E_{2|4,3} \\
E_{3|4,1} & E_{3|4,2} & E_{3|4,3}
\end{bmatrix},
\end{equation}
has full rank.\vspace{-1pt}

\section{Conclusions}\vspace{-5pt}
We have proven that as conjectured by Ac\'{\i}n {\em et al.}\ in Ref.\ \cite{Acin2016}, the maximal quantum violation of the elegant Bell inequality can be used to certify, in a device-independent way, two bits of randomness from one ebit. This demonstrates how fundamental tools in quantum information, namely, an ebit, a complete set of qubit MUBs, and the elements of qubit SIC-POVMs, are connected to maximal randomness.
An open question is whether a certification similar to ours would be possible with fewer measurement settings.
If not, this would sharpen the elegance of the protocol and strengthen the surprising connection between complete sets of MUBs and SIC-POVM elements, on one side, and optimal maximal randomness from maximal entanglement, on the other.

Concerning the practical aspects of randomness generation, it should be mentioned that violating different Bell inequalities is not equally costly in terms of statistics \cite{Peres2000,Gill2014}. 
Moreover, to certify device-independent generation of more that one random bit from an ebit,
it is often better to use a three-outcome POVM rather than a four-outcome POVM since the former is generally more robust against imperfections in the experimental setup \cite{Gomez2017}.

\begin{acknowledgments}\vspace{-5pt}
We thank Ingemar Bengtsson for fruitful discussions and for proposing improvements to the text.
We also thank Gustavo Ca\~{n}as for his help with Fig.\ \ref{Fig1}. A.C.\ acknowledges support from Project No.\ FIS2014-60843-P, ``Advanced Quantum Information'' (MINECO, Spain), with FEDER funds,
the FQXi Large Grant ``The Observer Observed: A Bayesian Route to the Reconstruction of Quantum Theory,''
and the project ``Photonic Quantum Information'' (Knut and Alice Wallenberg Foundation, Sweden).
\end{acknowledgments}

\end{document}